\documentclass[12pt]{article}
\usepackage[centertags]{amsmath}

\def\de{\delta}
\def\De{\Delta}

\def\tens{\otimes}

\def\de{\delta}
\def\De{\Delta}

\def\tens{\otimes}

\begin{document}

\title {Algebraic structure of $n$-body systems}

\author{${E.Sorace}^{1,2}$}

\maketitle

\centerline{{\small  ${ }^1$ INFN Sezione di Firenze}}
\centerline{{\small ${ }^2$ Dipartimento di Fisica, Universit\`a di
Firenze, Italy. }}

                \bigskip
\bigskip
\begin{abstract}
\noindent A general method to easily build global and relative
operators for any number $n$ of elementary systems if they are
defined for $2$ is presented. It is based on properties of the
morphisms valued in the tensor products of algebras of the
kinematics and it allows also the generalization to any $n$ of
relations demonstrated for two. The coalgebra structures play a
peculiar role in the explicit constructions. Three examples are
presented concerning the Galilei, Poincar\'e and deformed Galilei
algebras.
\end{abstract}

\medskip
\noindent
{\bf PACS}: 02.20 Sw, 02.20 Uw, 11.30 Cp, 03.65.Fd

\bigskip
\bigskip
\thispagestyle{empty}

\section{Introduction}
\bigskip

It has been recently found that the renormalization procedure in
Quantum Field Theory is intrinsically determined by an Hopf
algebra whose essential constituent has been soon recognized to be
the set of the parameters of the classical group of Virasoro with
their law of composition \cite{1}. The presence and the  utility of basic
algebraic concepts even in such an
elementary problem as the search of the so called ``relative
variables'' in classical and quantum mechanics will be illustrated
here. In this note indeed we introduce a method, whose use is based on the
coalgebraic structure of canonical commutation relations, which
allows for the explicit construction of the collective, {\it i.e.}
global and relative, canonical operators for any number $n$ of
elementary systems in a given kinematics once one has been able to
operate this transformation for $n=2$. It is also shown that there
are  classes of relations between single system operators and the
collective ones that once they hold for two of them, then they are straightforward
extended to any $n$. This is done by using the same algorithm
which generates the transformation of the operators. The
construction is {\it a priori} possible in any situation in which
operators are constructed in terms of single algebra generators
and the separation of the ``global operators'' is necessary. The
presentation is self explanatory, very euristic and
constructive. We tacitly suppose the existence of any object
necessary for the results.

The examples are
thus essential, not only to illustrate the physical utility of the method
but also to show that it is rather flexible and not mathematically void.
In section {\bf1} we give the general definitions and results. In
sections {\bf2}, {\bf3} we present examples from usual Galilei
and Poincar\'e kinematics
while useful applications are devised for a quantum algebra too in {\bf4}.
We use always $1d$ algebras,
thus avoiding the rotations whose consideration is not essential in
exemplifying the method. In {\bf5} there are some concluding remarks.

\bigskip

\section{Collective operators}

Let us consider an algebra $A$, whose elements can be
represented as hermitian operators on an Hilbert space, endowed with algebra morphism ${\Gamma}:
A\rightarrow A \tens A $,
\begin{eqnarray}
&&\Gamma a = \Sigma (a_{1})_{i} \tens (a_{2})_{i} ~~ \forall a
\epsilon A,~~~(a_{1})_{i},(a_{2})_{i}\epsilon A,\nonumber \\
&&\Gamma (ab) = \Gamma a \Gamma b,~~~\forall a,b \epsilon A, \label{za}
\end{eqnarray}

With such a morphism $\Gamma$ one can combine the two operators of
the single systems in {\it collective} ones, we call {\it global}.
The set of operators acting on the same space must be completed to
conserve the number of the operators of the original set of the
single systems, (such sets include too the Poisson algebra of
functions on a symplectic manifold).

This involves the introduction of the aforementioned {\it relative}
operators
which complement the global ones in the set of the
collective operators.

We thus introduce, by assuming it exists, another homomorphic
map $\de$ on $ A $:
\begin{equation}
\delta: A\rightarrow A \tens A ,\qquad
\delta (ab) = \delta a \delta b,~~~~~~\forall a,b \epsilon A,\qquad
\Gamma A \oplus \delta A = A\tens 1 \oplus 1\tens A  \label{ub}
\end{equation}

Moreover we impose the commutation property
\begin{equation}
\Gamma a  \delta b -\delta b \Gamma a \doteq [ \Gamma a,\delta
b]=0~~~~~~ \forall a,b \epsilon A    \label{uc}
\end{equation}

\noindent so that $\Gamma$ and $\de$ implement exactly a
transformation we may call canonical. If the generators of $A$
satisfy canonical relations we may call canonical the operators
recovered by the transformation.

As a direct consequence of the more general result below it is possible to
produce all the collective canonical operators for  $n$-body by
using only $\Gamma$ and $\de$ and the right tensor multiplication
$\tens 1$.

Indeed let us given $n$ morphisms, $\Gamma_j:A \rightarrow A \tens
A, (j=0,..n-1)$, not necessarily different, and a morphism $\de$
satisfying (\ref {ub}) and (\ref {uc}), with $\Gamma = \Gamma_j,~~
\forall j \epsilon (0,n-1)$.

Let us now consider the following expressions:

\begin{eqnarray}
&&\de a \tens  1^{{\tens}^{(n-2)}} ,\quad \nonumber \\
&&{\Gamma}^{(n-1)}\de a \tens  1^{{\tens}^{(n-3)}},\quad \nonumber \\
&&{\Gamma}^{(n-1)}{\Gamma}^{(n-2)} \de a \tens 1^{{\tens}^{(n-4)}} ,\quad \nonumber \\
&&...............................................................................\nonumber \\
&&{\Gamma}^{(n-1)}.......................{\Gamma}^{(1)}\de a ,\quad \nonumber \\
&&{\Gamma}^{(n-1)} ....................{\Gamma}^{(1)}
{\Gamma}^{(0)} a   \label{ad}
\end{eqnarray}

\noindent where the following notations have been introduced:

${\Gamma}^{(0)}=\Gamma_{\sigma
(0)},~~{\Gamma}^{(j)}=\Gamma_{\sigma(j)} \tens id^{{\tens}^{(j)}}$
with $x^{{\tens}^{(j)}}\equiv x\tens...x\tens x,~(j~~\text
{factors}~~x)$,  and $\sigma$ is any permutation of $i,
(i=0,..n-1)$.

The following {\it{proposition}} then holds:

"Each one of the previous expressions realizes the algebra $A$ in
the n-fold tensor product $A^{{\tens}^{(n)}}$.  Moreover the n
realizations are all mutually commuting."

The demonstration is by recurrence. So let us suppose the
proposition is true for n and apply to the left of  each one of
the n previous expressions {\ref{ad}}  the map ${\Gamma}^{(n)}$.
This map is a morphism ${A}^{{\tens}^{(n)}}\rightarrow
{A}^{{\tens}^{(n+1)}}$ so that the new expressions satisfy the
same algebraic relations as the previous ones. Let us now complete
the set with the new expression $\de a \tens 1^{{\tens}^{(n-1)}}$:
to end the proof it must be shown that it commutes with the $n$
new expressions. By construction the last ones are $n$-multilinear
sums of elements of the form $(\Gamma_{\sigma(n)}a_{(1)})\tens
a_{(2)}....\tens a_{(n)},~a_{(k)} \epsilon A,~ k \epsilon (1,n)$.

The commutators to evaluate are sums of elements:

$$[(\Gamma_{\sigma(n)}a_{(1)}) \tens a_{(2)}....\tens a_{(n)} ,\de a \tens
1^{{\tens}^{(n-1)}}]= [\Gamma_{\sigma(n)}a_{(1)},\de a]  \tens
a_{(2)}....\tens a_{(n)} = 0.$$

The proposition being true for $n=2$ is thus demonstrated  for any
n.

If one deals, as it is very probable, with only one morphism $\Gamma = \Gamma_j,~~\forall j\epsilon (0,n-1)$  then the
(\ref{ad}) give a straightforward procedure to generate the n-body expressions from the two-body ones.

By the way  in this case the proposition holds even if we change everywhere in (\ref{ad}) $\Gamma$ with $\de$.

Let us show now a general implication of the use of algorithmic definitions in generating relations
for $n$-body once they hold for $2$.

So suppose that for $2$ peculiar elements $a,b~ \epsilon A$ ($a$ may be
equal or not to $b$) it exists a
function $R$ on $A \tens A \times A \tens A \times A \tens A$  with
values in $A \tens A $, which can be extended
on  the direct product of growing tensor  powers of $A$, which makes
explicit a relation between
global and the relative operators built on $a$ and $b$  in the
form, {\it e.g.}:
\begin{equation}
{\Gamma}^{(0)}a = R(a_1, a_2, \de b)  \label{ae}
\end{equation}

where we have introduced the notation:
$$ z_{i}= 1\otimes 1\otimes 1....\otimes z\otimes
1\otimes......\otimes 1, $$ with $z$ acting on the i-th space.

Let us now apply to both sides of (\ref{ae}) one time the right
multiplication $\tens 1$ and another time the operation from the
left ${\Gamma}^{(1)}$. We thus get by exploiting identities
like  $f(a\tens 1)= f(a)\tens1$, with $a, f(a) \epsilon
A^{{\tens}^{(j)}},~~j~integer$ :
$$
{\Gamma}^{(0)}a \tens 1 = R(a_1, a_2, \de b \tens 1)\\,\\~~~~~~~~~
{\Gamma}^{(1)}{\Gamma}^{(0)}a = R({\Gamma}^{(0)} a \tens 1, a_3, {\Gamma}^{(1)} \de b)$$

\noindent where now $a_1, a_2$ must be read as $a\tens 1 \tens 1,
1\tens a \tens 1$ and R is valued in $A \tens A \tens A$. The
explicit relation between the $3$-body collective operators and
the $3$ single body ones is therefore:

$${\Gamma}^{(1)} {\Gamma}^{(0)}a = R( R(a_1, a_2, \de b \tens 1), a_3, {\Gamma}^{(1)} \de b)$$
It is now straightforward to find, by iterating $n-2$ times the
two previous operations, that for the $n$-body the following
implicit relation  between all the single and all the  collective
operators derived from $a$ and $b$ (the tensor product domain and codomain of R
is extended at each step) holds:

\begin{equation}
{\Gamma}^{(n-2)}...{\Gamma}^{(1)} {\Gamma}^{(0)}a =
R({\Gamma}^{(n-3)}...{\Gamma}^{(1)}{\Gamma}^{(0)} a \tens 1,
a_n,{\Gamma}^{(n-2)}...{\Gamma}^{(1)}\de b),   \label{af}
\end{equation}
and the general solution in the computable form of a recursive function is:
\begin{eqnarray}
&&{\Gamma}^{(n-2)}...{\Gamma}^{(1)} {\Gamma}^{(0)}a=\nonumber \\
&&R(R(..(R(a_1, a_2, \de b \tens 1^{{\tens}^{(n-2)}}),
a_3,{\Gamma}^{(1)}\de b \tens 1^{{\tens}^{(n-3)}}),..),a_n,
{\Gamma}^{(n-2)}..{\Gamma}^{(1)}\de b)  \label{kr}
\end{eqnarray}

\noindent which eventually recovers an explicit expression by
taking into account the concrete form of the $2$-body initial
relation (\ref{ae}). Let us remark now that $a$ could be any
expressions of the generators so that there can be interesting
cases in which $R$ is simply a primitive recursive function.
Moreover the demonstration deals only with the elements $a, b$ and
those expressions derived from them by using $\Gamma, \de$ so that
(\ref{kr}) could hold even if $\Gamma, \de$ don't fulfil their
defining properties on all $A$.

 As is well known  a morphism $\De A \rightarrow A\tens A$ is called {\it coproduct} when the
 {\it{coassociativity}} holds:
\begin{equation}
(\De \tens id)\De a = (id \tens \De)\De a, ~~ \forall a \epsilon A
\label{zb}
\end{equation}
In this case $A$ is a {\it coalgebra}, all the Lie and quantum
algebras stay in this category. A property to notice in this
context is that owing to the coassociativity (\ref{zb}) the action
of $\De$ can be univocally iterated to any $ A^{{\tens}^{(n)}}$
There are also algebras where the coassociativity is fulfilled
only modulo some equivalence: the {\it quasi-coalgebras}, and the
quasi-coassociative morphism is the {\it quasi-coproduct}. When we
deal with n representations ({\it rps} from now on) of $A$ we can,
by means of the map $\De$ and $id$ recover a set of {\it global}
operators on the product space satisfying exactly the original
algebra of the single components $L_a$, independently of the order
of $\De$ and $id$. In any Lie algebra the coproduct simply  reads
in algebraic terms:
$$\De L_a= L_a \tens  1 + 1 \tens L_a $$
When there is a basis of an algebra $A$ in which $\De$ gets this
form it is called a {\it primitive} coproduct. If $\De$ is
invariant after the interchange of the two base spaces in the
tensor product it is called {\it cocommutative}, any element built
in terms of the generators of a Lie algebra clearly shares this
property.  The ``barycenter formulas'' of the classical kinematics
are tied to the canonical coassociative coproduct. Thus the
starting point in the research of the collective operators must
be, if it exists, the coproduct. But one cannot find  in general a
$\de$ satisfying (\ref{uc}) with $\Gamma = \De$; clear examples
are given by semisimple Lie algebras. A near solution to this
problem could exist{\it e.g.} for nonsemisimple Lie algebras with
non null first class of cohomology, where an arbitrary scalar
variation can be given to the action of the global morphism on
some generators and a quasi-coassociative $\Gamma$ results.
Sometimes we can thus satisfy (\ref{uc}), at the price however of
the non univocity of some global operators. Actually to proceed
with $\Gamma \not= \De$ seems physically reliable only when one
deals with global operators of no direct physical meaning. Anyway
it must be remarked again that the collective set is completely
defined once the morphisms $\Gamma_j,  \delta$, whichever they
are, have been done. This will be illustrated  from the three
examples we present in the next sections which  share different
degrees of complexity.

\bigskip

\section{The Galilei Algebra}

Despite its ubiquitous presence in the contemporary Physics,
 as symmetry of the non relativistic  Q.M., the literature on the
Galilei group is not huge, and even in
general presentations \cite{2} the space devoted to collective
coordinates is  not large. Moreover in last times physical results
and researches mainly concerned the classical and quantum
statistical mechanics and the field theory implications of the
Galilean invariance \cite{3,4,5,6}. Therefore it maybe that an
extensive treatment of the collective position  operators in
$2$-body Galilei kinematics must be searched yet in \cite{7}. Thus
it will be instructive to apply firstly our method to the $1d$
Galilei group. The mass is chosen to be a Lie generator, this
implies the use of non projective representations with the
advantages that the Galilean symmetry is seen  from the physicist
viewpoint, see {\it e.g.} \cite{8}, and that this is the form
necessary to obtain the deformed version \cite {9}.

We start thus with the $3$ generators Lie algebra $gh(1)$ :
\begin{equation}
[B, P]= i M,~~~~~~~~[M, B] = [M, P]= 0;    \label{ba}
\end{equation}
It is the algebra of the purely spatial $1d$ extended Galilean
transformations where
B is the boost, P the momentum and the central generator M is the mass.

If one defines, by exploiting the localization with respect to the
center, the position generator $X=B/M$ one gets:
\begin{equation}
[X, P]= i 1,~~~~~~~   [M, X] = [M, P]= 0;    \label{bb}
\end{equation}
where 1 is the identity element of the enveloping algebra $U(gh(1))$.

The first commutator of (\ref{bb}) define a couple of Heisenberg
canonical operators. But the Lie coalgebraic structure in the
Heisenberg commutator is not compatible with X primitive if P is
primitive owing to ${\De}1 \doteq 1 \tens 1$. Indeed once the
momenta have been summed the corresponding positions must be
linearly combined with arbitrary coefficients whose sum is 1.

This is recovered by exploiting the algebraic status of M. In fact in the Lie algebra
(\ref{ba}) the coproduct amounts simply to:
\begin{equation}
\De P= P_{1}+P_{2},~~~~~~ \De B=B_{1}+B_{2},~~~~~~ \De M=M_{1}+M_{2}\label{bc}
\end{equation}

Consequently one has for X:
\begin{equation}
\De X= \De B/{\De M} = (M_{1}X_{1}
+X_{2}M_{2})/(M_{1}+M_{2})\label{bd}
\end{equation}
It is therefore very sensible to think in this case  to the Heisenberg
canonical set as a coalgebra with three generators.

A good well known map $\de$ is given by
\begin{eqnarray}
&&\de P = (P\tens M - M\tens P)/(\De M) ={\frac
{P_{1}M_{2}-M_{1}P_{2}}{M_{1}+M_{2}}}, \nonumber \\
&&\de X= X\tens 1- 1\tens X = X_{1}-X_{2}, \nonumber \\
&&\de M= M\tens M/(\De M)={\frac{M_{1}M_{2}}{M_{1}+M_{2}}}, \label{bf}
\end{eqnarray}

The expressions (\ref{ad}), with $\Gamma_j = \De$, are then the usual canonical Jacobi coordinates.

Anyway the algebra (\ref{ba})is a sub-algebra of the full Galilei Lie
algebra  $g$ one gets by adding a fourth Lie generator $E$, the energy,
whose non zero commutator is:
$$[B, E] = i P $$
The center  of $g$ is generated besides $M$ even by the quadratic
Casimir $ C=2 M E-P^2$. It is now obvious that the coproduct of
the energy cannot commute with all the relative operators. But the
generator $E$ cannot be written as a commutator and we can put
$\Gamma E \neq \De E$ Thus the set of collective operators can be
completed by introducing a {\it global energy} $\Gamma E$ and a
{\it relative energy} $\de E$. They can be found by imposing that:
$2 \Gamma M~ \Gamma E - \Gamma P^2$ and $2 \de M  ~\de E - \de
P^2$ are Casimir. The result is:
\begin{equation}
\Gamma E = {\frac{M_1 E_1 +M_2 E_2+P_1 P_2}{M_1+M_2}} \label{jg}
\end{equation}
and
\begin{equation}
\de E =(M\tens E + E \tens M)- P \tens P)/(\De M)= {\frac{M_1 E_2 +M_2
E_1-P_1 P_2}{M_1+M_2}}= E_1+E_2-\Gamma E  \label{bg}
\end{equation}

The definition of
$\Gamma E $ is anyway coassociative modulo global Galilei invariant operators.

The application of (\ref{ad}) gives the expressions for any $n$.
Let us notice that (\ref{bg}) is in a
form where the
general recursive formula (\ref{kr})  is trivially explicited so that
we have for any $n$:
\begin{equation}
\Sigma E_j ={\Gamma}^{(n-2)}...{\Gamma}^{(0)}E
+{\Gamma}^{(n-3)}...{\Gamma}^{(1)}\de E\tens 1+...+{\Gamma}^{(1)} \de E\tens
 1^{{\tens}^{(n-3)}}+ \de E\tens 1^{{\tens}^{(n-2)}} \label{jh}
\end{equation}
Let us observe  also that by choosing $a = b = P^2/(2 M)$ and then
recovering from the $2$-body that $R(x,y,z)=x+y-z$ the formula
(\ref{kr}) gives immediately that the sum of the $n$ single
kinetic energies transforms in the identical formal expression in
terms of the collective Jacobi set. It must be remarked that the
map $\de$ (\ref{bf}), doesn't satisfy coassociativity nor it is a
{\it coaction}, ({\it i.e.}: $(\De \tens id)\de a$ is not equal to
$(id \tens \de) \de a, \forall a \epsilon A $). Indeed it is just
the initial support for the action of the globalizing and
injecting operations. It can be shown by direct calculation that
there is no coassociative $\de$ producing all the previous
properties in the Galilei algebra. Of course one can introduce
functions of the masses as factors in the definition of  $\de $,
as, {\it e.g.}, in the analysis of the $ 1d$ integrable many-body
Schr\"odinger equation by McGuire \cite{10}. A $\de$ actually
quasi-coassociative can be obtained in this way, with a lack of
completeness however as relative and global masses happen to be
the same.

It is worth noticing that the analogous coproduct and the same
role of the mass hold in the three-dimensional situation, where
the Heisenberg set can be derived again by a sub-algebra of the
extended Galilei. In this case the expressions of the $2$-body
collective operators can be much more composite, following the
dynamical problems one has to face. But, as shown before, once the
collective expressions have been found for $2$ the algorithm to
give expressions for $n$ is straightforward.

 \bigskip

\section{The Poincar\'e Algebra }

The proposals about the localization and the canonical operators of
the position in special
relativity are not univocal, see {\it e.g.} \cite{11} and references therein.
We adopt here the one,
firstly studied in \cite{7}, based on the Weyl algebra,
analyzed and exploited in \cite{12} where
the hamiltonian dynamics
of $1$ and $2$ scalar or spinning relativistic particles was written.
Coulomb and Schwartzschwild type $2$-body interactions were
covariantly introduced in the mass square and the dynamics
of two scalar particles completely solved, with results in very good
agreement with field calculations (see \cite{13} also). Some very
encouraging quantistic
estimates were also done for $2$ and $3$ interacting scalar particles
 \cite{14}.
Moreover operators with identical expressions, although there the Weyl
algebra is included in the conformal one, have been independently
rediscovered and
proposed as the quantum observables of
relativistic spinning particles in many recent papers see \cite{15}
and references therein.
We exploit the cohomological based possibility of adding a
2-body Weyl invariant operator to the global dilatator
defined by the primitive coproduct. Thus the coassociativity holds
only modulo Weyl invariant operators and the global operators
involving the dilatators are strictly dependent on the order of the
$\Gamma$ and $id$.

We discuss now the $(1,1)d$ situation.
The analogous in the Poincar\'e kinematics of the $(2.2)$ is given by
the $E(1,1)$ Lie  algebra :
$$[B,P]=i E, ~~~~~~~~~~~~[B,E]= i P,~~~~~~~~~~~ [E,P]=0.$$
However, to get a time operator,  the starting point of our
procedure must be the Weyl algebra, obtained by adding as fourth
Lie generator the dilatator $D$:
$$ [D,P]=-i P,~~~~~~~[D,E]= -i E,~~~~~~~~~ [D,B]= 0 $$
We construct then two commuting Heisenberg pairs  by defining the two
``Lorentz (1,1)-vectors'': $(P,E)$ and $(X,T)$,
\begin{equation}
 X= M^{-2} (D P+ B E), ~~~~~~~T = M^{-2} (D E+ BP)
\label{ag}
\end{equation}
where $M^2=E^2-P^2$ is the Casimir of $E(1,1)$ and
one has $[X,P]= i,~~[T , E]= -i$, all the other
commutators being zero. Those operators are the building blocks of the
1-body.
 It must be remarked however that the dynamics of such
systems must be generated by Hamiltonians conserving the
Poincar\'e invariant mass and that the maximal invariance can be
the Poincar\'e symmetry, not the Weyl one, because in any
situation the physical time $T $ at least must change with any
evolution parameter: all that is done in quite natural manner in
this framework. The projection on the irreducible rps of $E(1,1)$
is indeed the equivalent of the classical reduction procedure on
the fixed mass sub-variety. Let us now discuss the $2$-body
collective scheme. It reads

$\Gamma E=E_{1}+E_{2},~~~~~\Gamma P=P_{1}+P_{2},~~~~~
\Gamma B=B_{1}+B_{2} $

\noindent and

$\Gamma D=D_{1}+D_{2}+ c$

\noindent where $c$ is an arbitrary element in the center of the
global Weyl in the tensor product, allowed because D never appears
on the right member of the commutations relations  (this happens
in the $(3,1)d$ case also). We have thus:

$\Gamma M= ((\Gamma  E)^2-(\Gamma P)^2)^{1/2}$ and the
``quasi-coproduct'' of $X, T $ is given by
$$
\Gamma X= (\Gamma M)^{-2} (((\mu_{1})^2  X_{1} +(\mu_{2})^2
X_{2})- t (P_{1}E_{2}-P_{2} E_{1})+ (2 i
+c)(P_{1}+P_{2}))$$
$$\Gamma T = (\Gamma M)^{-2} (((\mu_{1})^2 T_{1}+(\mu_{2})^2 T_2)+ r
(P_{1} E_{2}-P_{2} E_{1})+ (2 i
+c)(E_{1}+E_{2}))$$
where $(\mu _{A})^2= (E_{A})^2-(P_{A})^2+(E_{1} E_{2} -
P_{1} P_{2})$ so that $(\mu _{1})^2+(\mu _{2})^2)=(\Gamma M)^2$, and
it is

$ r= X_{1}-X_{2},~~~ t=T_{1}-T_{2}$.

$ q= (P_{1}-P_{2})/2,~~~ u=(E_{1}-E_{2})/2$

Let us choose
$c= -2 i -(u t-q r)$: it is
straightforward to show that $(\Gamma \tens id)\Gamma D - (id \tens \Gamma)\Gamma
D$ is again an operator invariant under the global $3$-body Weyl
algebra. A good set of relative
operators is then obtained by adding the definitions

$\de X= \tilde r=(\Gamma E~ r - \Gamma P~ t)/(\Gamma
M),~~~~
\de P=\tilde q =(\Gamma E~ q-\Gamma P~ u)/(\Gamma M)$

$\de T =\bar r=(\Gamma E~ t -\Gamma P~ r)/(\Gamma M),~~~~\de E= \bar q=(\Gamma E~ u -\Gamma P~ q)/(\Gamma M)  $

Together with $ \Gamma X,~ \Gamma T $, and $\Gamma P,~\Gamma E$
they give a complete set of canonical and ``covariant''(invariant
in this $1d$ case) operators as a direct calculation can confirm.
The relevant property of this set is the existence of a relation:
\begin{equation}
(\Gamma M)^2=(((M_{1})^2+(\tilde q)^2)^{1/2}+((M_{2})^2+(\tilde q)^2)^{1/2})^2
\label{an}
\end{equation}
recovered by eliminating $\de E$ from the collective expressions
of $(M_1)^2$ and $(M_{2})^2$ and solving in $(\Gamma M)^2$. By
projecting on definite values $(M_1)^2=m_{1}^2,
~(M_{2})^2=m_{2}^2$ one recovers for the relativistic $2$-body a
rigorous hamiltonian formulation in terms of one global time,
while the relative time $\de T =\bar r$ is ignorable and can be
chosen {\it a posteriori} to reconstruct the dynamics in the
higher dimension. At this point one can introduce interactions
depending on $|\tilde r|$. Clearly the physical description is
given at this level, the galilean limit too must be checked there.

It is now possible to extend straightforward (\ref{an}) to any number
of massive
Poincar\'e representations because it is given explicitly in the form of
relation (\ref{ae}). The absence of angular momenta in
those $(1,1)d$ models avoid any problem of formal covariance (as opposed to
the commutativity of the components of the position, see
\cite{12}). It is thus possible to
construct recursively, by adopting the formulas (\ref{ad}) for the
$n$-body  and the
corresponding expressions (\ref{kr}) with nested square roots, a
genuine relativistic hamiltonian system of
$n$ interacting particles, with $n$ given masses and one global
physical time.

\bigskip

\section{The Quantum framework}

The definitions (\ref{ad}), (\ref{af}) depend on a canonical map
and thus they can be in principle applied to any coalgebra. The
crucial problem is to find a good map $\de$ for the $2$-body
system. It is thus interesting to analyze from this view point the
operators of the quantum version of the Galilei algebra \cite{9},
where (\ref{uc}) cannot be completely realized. This deformed
algebra has found physical applications directly as kinematical
symmetry of many-body quantum dynamics on lattice \cite{9}.
Moreover its unitary irreducible rps have been studied by inducing
on the non commutative space of parameters and they appear in
agreement with those of Heisenberg on the lattice, but the
recovering of unitary irreducible rps in the usual way in the
common space of  the product of two is rather problematic,
notwithstanding the algebra has a real form although rather
unconventional \cite{17}.

Thus let us introduce the coalgebra $gh_{a}(1)$  having the same
$3$ generators and algebraic relations as $gh(1)$ and non trivial
coproduct of B and M:
\begin{eqnarray}
&&\De P= P\tens 1+1\tens P  \nonumber \\
&&\De B= B\tens exp(iaP)+exp(-iaP)\tens B \nonumber \\
&&\De M= M\tens exp(iaP)+exp(-iaP) \tens M \label{tf}
\end{eqnarray}
where the {\it length} $a$ is the deformation parameter and having
defined again $X=B/M$ one gets
\begin{equation}
\De X ={\frac {\De B}{\De M}}=
{\frac{M_1 X_1+M_2 X_2
exp(-ia(P_{1}+P_2))}{M_1+M_2 exp(-ia(P_1+P_2))}}
\end{equation}
We take obviously $\Gamma = \De$ but
let us observe that $\De M$ is a Casimir of the algebra $\De A$
but it is not a central element of $A \tens A$. Its expression implies
that it is
impossible to get $\de$ such that both $ \de X$ and $ \de P$
commute with $\De X$, $\De P$ and even with $\De M$.
The map $\delta$ we define is the following:
\begin{eqnarray}
&& \delta X = X_1 - X_2  \nonumber \\
&&\delta P ={\frac{i}{a}} log({\frac{\Gamma M}{(M_1+M_2)}}) \nonumber \\
&&\delta M ={\frac{ M_1 M_2}{(M_1+M_2)}} \label{al}
\end{eqnarray}
and we have two couple of commuting canonical operators, although
not a direct product of the two triples. Indeed there is a deformed
commutator:
$$ [\delta X, \Gamma M]= a \Gamma M   \,$$
By looking at the structure of the expressions (\ref{ad}) one sees
that in this case they produce n distinct realizations of the
algebra which however are not commuting between them. It must be
remarked again that $\De M,~ \De X$ are given {\it a priori} and
$\de X$ has the form necessary to commute with total momentum
while the remaining expressions have correct relations. Thus the
previous choice must be accepted and one has to pay the price of a
deformation of canonicity, starting from $n=3$, in the collective
formulation.

A quasi-associative energy $E$ completes the Galilean deformed
algebra. The resulting nonstandard
$$[B,E]=(i/a)sin(a P)$$
determines a Casimir $C= M E- (1/a^2)(1-cos(aP))$, from which we
define the deformed kinetic energy:
$$T=(1/(M a^2))(1-cos(aP))$$
It is then straightforward to obtain for the $2$-body operators:
\begin{eqnarray}
&&T_{1}+T_{2}=(1/(\De M~a^2))(1-cos(a \De P))+(1/(\de
M~a^2))(1-cos(a~\de P))= \nonumber \\
&&\De T + \de T  \label{ar}
\end{eqnarray}
We are again in a situation where an explicit elementary
expression of the (\ref{af}) exists and the previous anomalies
cannot affect the result given by (\ref{kr}). Indeed we are using
only the  abelian coalgebra generated by $P$ and $M$, with their
coproducts. Therefore we can be sure of the existence of the set
of trigonometric identities which state in the deformed case the
same theorem about the kinetic energies as in the classical one:
\begin{equation}
\Sigma T_{j}={\De}^{(n-2)}...{\De}^{(0)}T
+{\De}^{(n-3)}...{\De}^{(1)}\de T\tens 1+...+{\De}^{(1)} \de T\tens
 1^{{\tens}^{(n-3)}}+ \de T\tens 1^{{\tens}^{(n-2)}}      \label{bl}
\end{equation}
This is the kinetic part of a lattice Hamiltonian. If one searches
for values of observables such that the kinetic energy is given
only by the barycenter term the result is that all the relative
momenta must be zero, {\it i.e.}:
\begin{eqnarray}
&&\De^{(0)}M=M_{1}+M_{2}\nonumber\\
&&{\De}^{(j)}...{\De}^{(1)}\De^{(0)}M={\De}^{(j-1)}...{\De}^{(1)}\De^{(0)}
M+M_{j+2},~~~~j\epsilon~(1, n-2).\label{b,m}
\end{eqnarray}
It has been demonstrated that when all the masses are equal the system
(\ref{b,m}) gives exactly the Bethe conditions for the
momenta of $n$-magnons bound states of the XXX model and the
right spectrum of the energy \cite{9}.

It is possible to introduce in the same way as in the classical case the global energy
and the relative one $\de E$:
$$\de E = {\frac{m_{1} E_{2}+m_{2} E_{1}}{m_{1}+m_{2}}}+ \de T -
{\frac{m_{1} T_{2}+m_{2} T_{1}}{m_{1}+m_{2}}}$$  whose non
deformed limit is (\ref{bg}). The global energy is $
E_{1}+E_{2}-\de E$, which - like $\De T$ - doesn't commute with
$\de X$. A sum rule formally identical to (\ref{jh}) can be
written however.

\bigskip

\section{Concluding remarks}

An intuitive method of constructing collective classical canonical
coordinates or quantum mechanical operators for $n$-body on the
ground of their expressions for $n=2$ has been precisely
formulated and demonstrated by means of algebra morphisms,
constructed on the basis of the coalgebra of the systems. Examples
from Galilei, Poincar\'e and deformed Galilei are discussed. An
interesting result is the ability of writing immediately for $n$
relations calculated for $2$. A further point worth to be studied
is the way to apply the algorithm in field theory and the possible
connection to the integrability suggested by section {\bf 4}.
Preliminary analysis of those problems are {\it in fieri}.

Concerning the coproduct it must be stressed that its possible
substitution by the morphism $\Gamma$ is  essential in allowing a
rigorous and physically good description of the many-body
relativistic systems in our approach to the Poincar\'e systems.
From this view point the inclusion of the Weyl in the larger
conformal algebra as in \cite{15} may generate problems, as in
that case there is no space to substitute the coproduct of $D$
with a morphism having $c \neq 0$. This remark leads us again to
enhance a very general point sometimes ignored in the practice,
owing to the long monopoly of the Lie primitive structures; {\it
i.e.} that a complete knowledge of an algebra can be obtained only
by the knowledge of the coalgebra too. All that is very important
in those attempts to grasp quantum gravity by means of
noncommutative geometries, implied {\it e.g.} by the introduction
of deformed relativistic kinematics, strongly supported in last
years by the preliminary astrophysical measures concerning
gamma-ray bursts and the possible violation of the GZK threshold
in cosmic rays, see \cite{17,18} and references therein.
    A deep analysis of the collective operators connected to the proposed
deformations of the Poincar\'e kinematics could then be very
useful in formulating their phenomenological implications. Indeed
one exotic relation of dispersion is in itself not enough, but if
it is accompanied by the emergence of $2$-body spectra deduced
from noncocommutative coalgebra it will be read as a clear
signature of a noncommutative space-time.
\bigskip

{\it Acknowledgments:}

\noindent I thanks M.Tarlini for very helpful discussion and
valuable criticisms.

\end{document}